\documentclass[journal]{IEEEtran}
\usepackage{graphicx}
\usepackage{xcolor}
\usepackage{amsmath}
\usepackage{amssymb}
\usepackage{authblk}
\usepackage{caption}
\usepackage[numbers,sort&compress]{natbib}
\usepackage{authblk}

\begin{document}

\title{High-Transmission Cryogenic Fiber-Optic U-Bench Filter for Ultra-Low-Background Single-Photon Detection in Superconducting Devices}

\author[1]{Elmeri O. Rivasto}
\author[2]{Katharina-Sophie Isleif}
\author[1]{Friederike Januschek}
\author[1]{Axel Lindner}
\author[3]{Manuel Meyer}
\author[3]{Joakim R. Rosenkilde}
\author[2]{Jos\'e Alejandro Rubiera Gimeno}

\affil[1]{Deutsches Elektronen Synchrotron DESY, Notkestr. 85, 22607 Hamburg, Germany}
\affil[2]{Helmut-Schmidt-Universität (HSU), Holstenhofweg 85, 22043 Hamburg, Germany}
\affil[3]{CP3-origins, University of Southern Denmark, Campusvej 55, 5230 Odense, Denmark}

\maketitle

\noindent
\begin{abstract}
\boldmath
We report on our progress in implementing a cryogenic fiber-optic U-bench hosting an ultra-narrow bandpass filter, designed to mitigate the dominant fiber-coupled 300\,K black-body background observed for superconducting single-photon detectors. While U-benches can be routinely aligned at room temperature to achieve minor optical losses, cooling them down results in substantial misalignment due to thermal contraction of their components. To mitigate the resulting losses, we have integrated cryo-compatible piezo stepper motors to the U-bench, enabling its \textit{in~situ} realignment while cold. In particular, this allows us to shift the temperature dependent transmission window of the used filter by rotating it with respect to the beam axis. First results show that our U-bench filter reaches a transmission coefficient of 38\% (4.2\,dB loss) at 40\,K temperature. We project that the U-bench filter can reduce the black-body background down to $10^{-8}$\,Hz and $10^{-3}$\,Hz ranges for 1064\,nm and 1550\,nm wavelengths, respectively. With the expected background rates and the herein reported transmission coefficient, applying the U-bench filter in series with a transition edge sensor shows great potential in increasing the sensitivity of dedicated rare-event searches. We project that the noise-equivalent power for 1550\,nm wavelength could be reduced down to $10^{-20}\,\mathrm{W}\,\mathrm{Hz}^{-1/2}$ range if intrinsic background sources can be sufficiently mitigated.
\unboldmath
\end{abstract}

\section{Introduction}
\label{introduction_section}
\noindent \IEEEPARstart{S}{uperconducting} single-photon detectors (SSPDs) form a broad class of devices characterized by high detection efficiencies and low background (dark count) rates \cite{Hadfield2009single, Eisaman2011invited, Morozov2021superconducting}. Energy resolving \textit{transition edge sensors} (TESs) \cite{De2024transition} are arguably one of the best performing class of SSPDs operating in the optical and near-infrared range. As their main advantage, the energy sensitive detection enables the effective mitigation of dark counts induced by intrinsic background sources, such as radioactive decays and cosmic rays \cite{Shah2022characterising, Shah2022TES, Meyer2024first, Manenti2024dark, Rivasto2026binary, Meyer2026characterization}. Subsequently, the extrinsic background dominated by fiber-coupled room temperature black-body radiation has been concluded as the limiting background source \cite{Miller2007superconducting, Yamashita2010temperature, Niwa2017few, Gimeno2024tes, Rubiera2025simulation, Meyer2026characterization}. The rate of the black-body background in the region of interest for a mono-energetic source is ultimately set by the energy resolution of the TES, determining its signal-to-noise ratio.

In this work, we demonstrate how a narrow bandpass pass optical filter operated at cryogenic temperatures in series with a TES would improve its energy selective photon detection (or enable energy-selective photon detection for energy-insensitive SSPDs) and subsequently provides drastic reduction in the black-body background (Section~\ref{black-body_background_section}). We then provide a technical description of our prototype (Section~\ref{Technical_Description_section}) and present first performance tests regarding mitigation of thermal contraction induced optical losses (Section~\ref{cooldown_section}).

\section{Black-body Background for TES}
\label{black-body_background_section}
\begin{figure}[t!]
    \centering
    \includegraphics[width=0.8\linewidth]{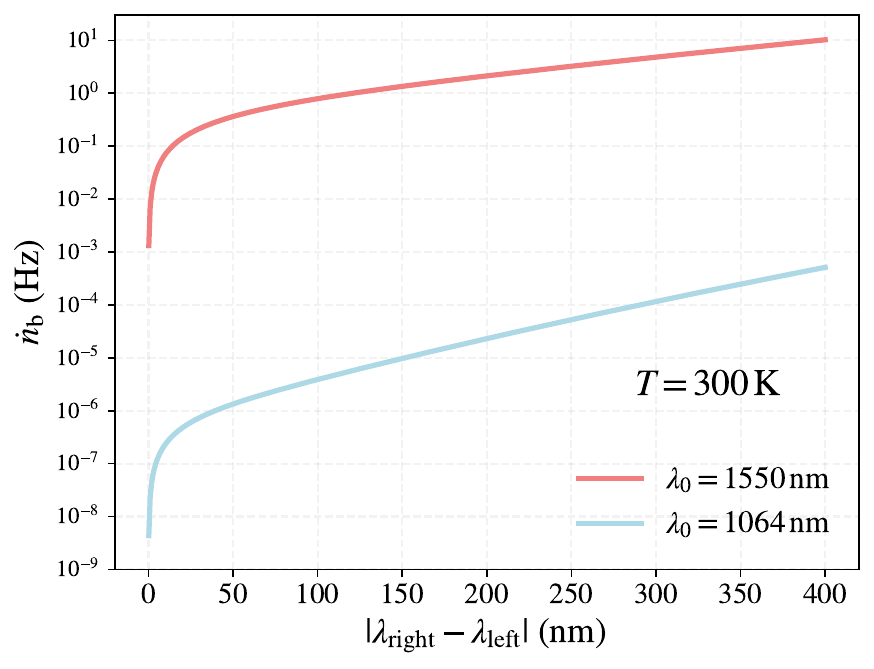}
    \caption{The 300\,K fiber-coupled black-body background rate as a function of the effective energy resolution of a TES calculated from Eq.~\eqref{Plancks_Spectrum_Eq}, assuming a fiber with $R_\mathrm{fib}=3\,$\textmu m and $\mathrm{NA}=0.14$ for both wavelengths. The horizontal axis represents the overall width of the integration region, i.e. $|\lambda_\mathrm{right}-\lambda_\mathrm{left}|=2\cdot n\Delta \lambda$. Equivalently, horizontal axis corresponds to the transmission window of a bandpass filter in front of the TES.}
    \label{Ndot-dlam_Fig}
\end{figure}

\noindent We have previously measured an energy resolution (full width at half maximum) of $\Delta E = 0.15\,\mathrm{eV}$ ($\Delta \lambda = 137\,\mathrm{nm}$) at $E_0=1.17\,\mathrm{eV}$ ($\lambda_0=1064\,\mathrm{nm}$) for one of our several TES modules, which we use in the calculations for the rest of the paper \cite{Gimeno2024tes} (this value is compatible with comparable TES modules \cite{Hummatov2023fast, Li2025tes}). For the fixed energy resolution, the black-body background rate ($\dot{n}_\mathrm{b}$) can be then estimated from Planck's spectrum by integrating over the wavelengths $\lambda \in [\lambda_0-n\cdot \Delta \lambda,\,\lambda_0+ n\cdot\Delta \lambda]$, where $n\in\ \mathbb{R}$ is an arbitrary factor affecting the analysis efficiency (which determines the detection efficiency, $\eta$). That is,
\begin{equation}
\label{Plancks_Spectrum_Eq}
    \dot{n}_\mathrm{b} = \int d\Omega \int dA \int_{\lambda_0-n\cdot \Delta \lambda}^{\lambda_0+n\cdot \Delta \lambda}  \frac{2c}{\lambda^4} \cdot \frac{1}{\mathrm{e}^{hc/\lambda k_\mathrm{B}T} -1} d \lambda,
\end{equation}
where $h$ is Planck's constant, $k_\mathrm{B}$ is Boltzmann's constant, $c$ is the speed of light and $T$ is the source temperature (here 300\,K) \cite{Schroeder2020introduction, Rubiera2025simulation}. For simplicity, we fix $n=3$ in order to assume near-unity detection efficiency at the central wavelength, and we further neglect wavelength dependent losses due to fiber transmission or fiber curling. The integrals over the solid angle~($\Omega$) and area~($A$) are determined by the fiber geometry where the black-body radiation is coupled into \cite{Rubiera2025simulation}. By definition the solid angle can be calculated as $\Omega = 2\pi\cdot (1 - \mathrm{cos}\,\theta)$, where $\theta$ is the acceptance angle of the fiber determined by its numerical aperture as $\theta=\mathrm{sin}^{-1}\,\mathrm{NA}$. The area integral can be simply evaluated based on the radius of the fiber core as $A=\pi R_\mathrm{fib}^2$. In the case of a typical single-mode fiber, with $R_\mathrm{fib}\approx 3\,$\textmu m and $\mathrm{NA=0.14}$, one obtains $\dot{n}_\mathrm{b} \sim  10^{-4}\,$Hz for $\lambda_0=1064\,$nm, coinciding with experimental observations \cite{Rubiera2025simulation, Rivasto2026binary}. Evidently, $\dot{n}_\mathrm{b}$ reduces exponentially as $\Delta \lambda$ is decreased, but the most dramatic change can be seen at very low values of $\Delta \lambda\lesssim 10\,$nm as illustrated in Fig.~\ref{Ndot-dlam_Fig} for both 1064\,nm and 1550\,nm wavelengths. While the energy resolution of a TES is typically well beyond $\Delta \lambda > 100\,$nm, such small $\Delta \lambda$ windows can only be achieved by applying a narrow bandpass filter in front of the TES. For experimentally plausible filter linewidth of $\Delta\lambda=0.5\,$nm, the fiber-coupled black-body backgrounds could be mitigated down to $\sim 10^{-8}\,$Hz and $\sim10^{-3}\,$Hz ranges for center wavelengths of 1064\,nm and 1550\,nm, respectively.  

\begin{figure}[t!]
    \centering
    \includegraphics[width=0.95\linewidth]{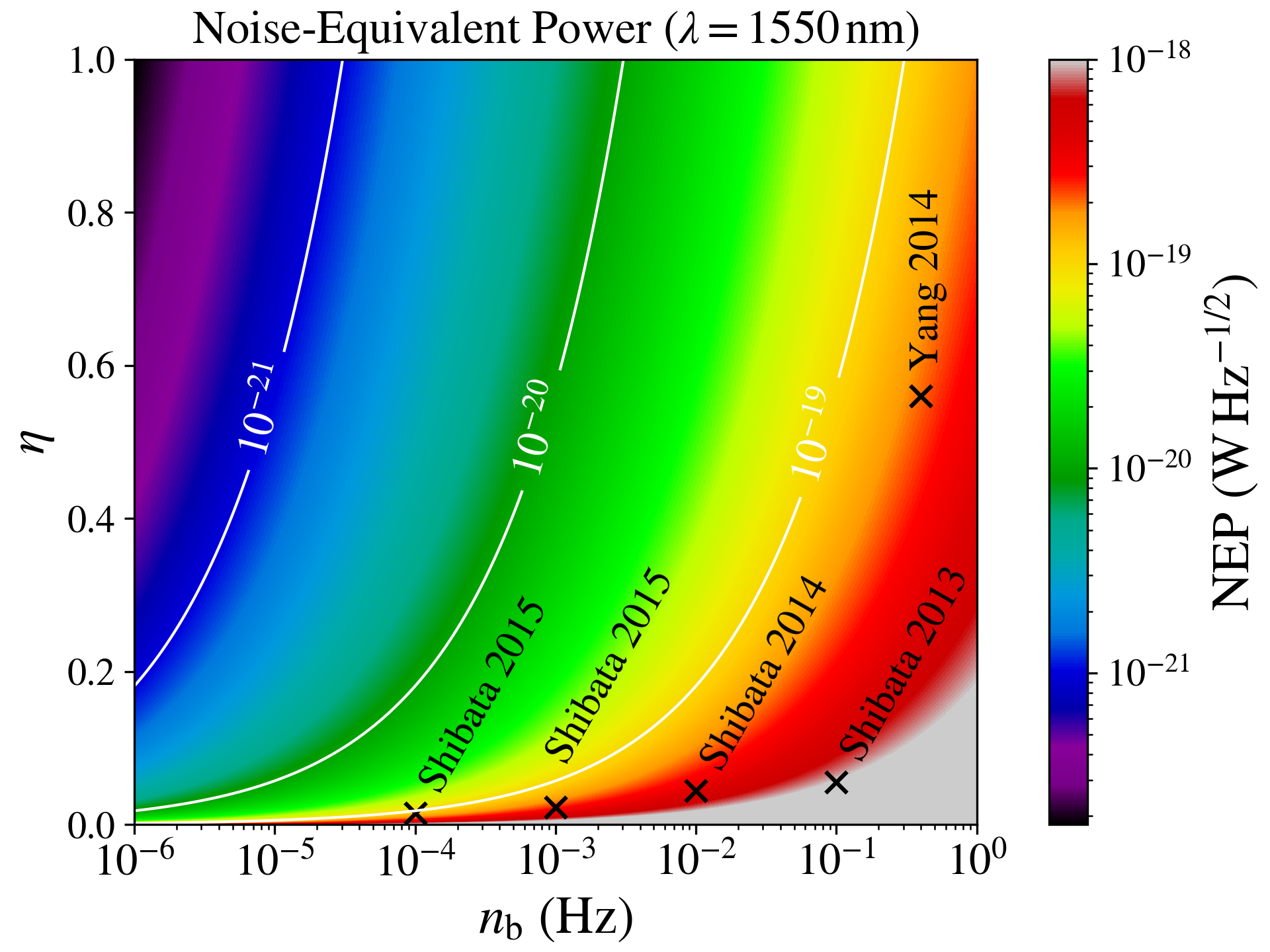}
    \caption{The noise-equivalent power (Eq.~\eqref{NEP_Eq}) as a function of dark count rate and detection efficiency for 1550\,nm photons. The figure displays the previously achieved NEPs for SSPDs using cryogenic optical filters implemented by Shibata~\textit{et al.} \cite{Shibata2013superconducting, Shibata2014quantum, Shibata2015ultimate} and Yang~\textit{et al.} \cite{Yang2014superconducting}.} 
    \label{NEP_and_S_Fig_ab}
\end{figure}

While the mitigation of black-body background by applying low temperature bandpass filters has been successfully demonstrated in the existing literature \cite{Shibata2013superconducting, Shibata2014quantum, Yang2014superconducting, Shibata2015ultimate, Shibata2016snspd}, it has always come at a cost of significant reduction in detection efficiency ($\eta$) as illustrated in Fig.~\ref{NEP_and_S_Fig_ab}. It should be noted that the quantum efficiency of TESs and superconducting nanowires can be close to unity \cite{Lita2008counting, Fukuda2011titanium, Hu2020detecting, Meyer2026characterization}, and thus we assume that the overall $\eta$ is (mostly) governed by the transmission coefficient of the applied filter. The required balance between the $\dot{n}_\mathrm{b}$ and $\eta$ ultimately depends on the application. The conventionally used figure of merit for SSPDs, in particular when applied in information transfer, is \textit{noise-equivalent power} (NEP) given by  \cite{Donati2021photodetectors, Hadfield2009single}
\begin{equation}
\label{NEP_Eq}
    \mathrm{NEP}= \frac{hc}{\lambda} \cdot  \frac{\sqrt{2 \dot{n}_\mathrm{b}}}{\eta} .
\end{equation} 
The dependence of $\mathrm{NEP}$ on $\dot{n}_\mathrm{b}$ and $\eta$ for 1550\,nm wavelength is illustrated in Fig.~\ref{NEP_and_S_Fig_ab}(a). One can see how the low $\eta$ reported for the filter implementation of Shibata \textit{et al.} \cite{Shibata2013superconducting, Shibata2014quantum, Shibata2015ultimate, Shibata2016snspd} limits their NEP well above $10^{-20}\,\mathrm{W}\,\mathrm{Hz}^{-1/2}$ range. Assuming that dark count rate would be limited to $10^{-3}\,$Hz, one could reach $\mathrm{NEP}=10^{-20}\,\mathrm{W}\,\mathrm{Hz}^{-1/2}$ by increasing the detection efficiency only up to $\eta \gtrsim 50\%$, which we consider readily achievable.

We presume that the optical losses in cryogenic U-bench filters, such as the ones reported by Shibata~\textit{et~al.}~\cite{Shibata2013superconducting, Shibata2014quantum, Shibata2015ultimate, Shibata2016snspd}, are due to their non-uniform thermal contraction during the cooldown from ambient environment. The only way to mitigate the resulting misalignment losses is to realign the system \textit{in~situ} in the cryogenic environment. For the rest of this paper, we present and discuss the operational performance and limitations of our custom build remotely adjustable fiber-optic U-bench filter at 40\,K temperature.

\section{Technical Description}
\label{Technical_Description_section}
\begin{figure*}
    \centering
    \includegraphics[width=0.95\linewidth]{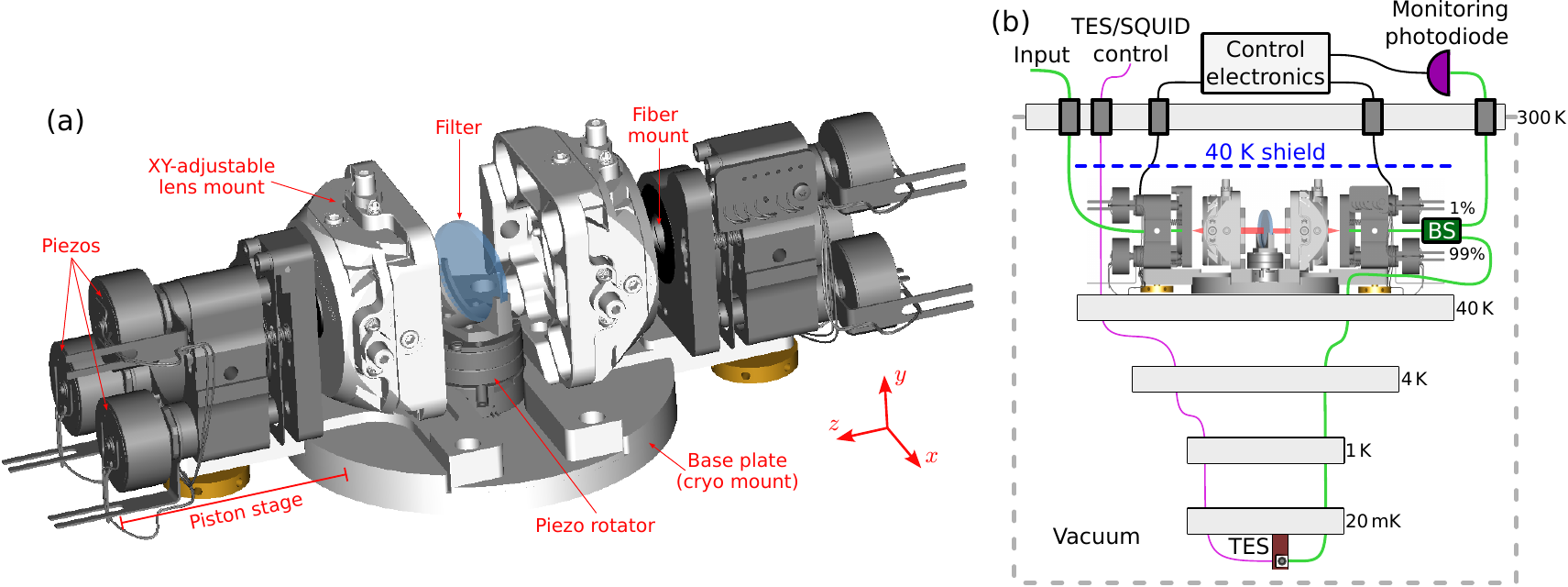}
    \caption{(a) A schematic illustration of the implemented fiber-optic U-bench filter. (b) The planned integration of the U-bench filter with the TES inside a dilution refrigerator.}
    \label{filter_bench_schematics_Fig_ab}
\end{figure*}
\noindent A schematic illustration of the implemented fiber-optics U-bench filter is shown in Fig.~\ref{filter_bench_schematics_Fig_ab}(a). A single mode optical fiber carrying 1064\,nm photons (HI1060), with mode field diameter of $6.2\,$\textmu m, is mounted on a Ti-based cryo-compatible piston stage (CTTPS1,~JPE) featuring three piezo stepper motors (CLA2201,~JPE). These motors enable \textit{in~situ} tip and tilt alignment of the fiber tip at $1$\,nm precision at cryogenic temperatures. The beam exiting the fiber is collimated using a monochromatic aspheric lens (60FC-4-A8-07, Schäfter+Kirchoff) having $f=8\,\mathrm{mm}$ focal length and 1050--1550\,nm working range. The collimating lens is mounted on a separate $xy$-positioner (POLARIS-05XY,~Thorlabs), enabling the manual adjustment of the lens position within the plane ($xy$-plane) perpendicular to the beam axis ($z$-axis) with a 212\,\textmu m/revolution sensitivity. We want to point out that the $xy$-positions of the lenses are currently the only degrees of freedom in our setup that cannot be controlled \textit{in~situ}. Meanwhile, the fiber~tip--lens distance (focus) can be adjusted by simultaneously moving the three piezos of the piston stage uniformly. The collimated beam then passes through a multi-cavity ultra-narrow bandpass filter (1065-1.8~OD6, Alluxa) centered at $\lambda_0 = 1065\,$nm with transmission window $\Delta \lambda < 2\,$nm (FWHM). The filter is mounted on top of a cryo-compatible rotating piezo stepper motor (CSR1,~JPE) enabling \textit{in situ} control of its angle of incidence with respect to the collimated beam, which we need for correcting the blue-shift of the filter's transmission window due to thermal contraction of its structure. The filtered beam is then coupled back to a single-mode fiber with a symmetrical setup described above. 

The above described setup was mounted on top of a custom-build Al/Invar base plate attached on top of the 40\,K plate of a dilution refrigerator (SD, Bluefors), inside which we operate our TES modules at 20\,mK base temperature. The planned setup with the U-bench filter integrated in series with the TES is illustrated in Fig.~\ref{filter_bench_schematics_Fig_ab}(b). However, in the initial experimental cooldowns of the U-bench filter presented below, its output was connected directly to a vacuum fiber-feedthrough. The piezos of the filter bench were manually controlled via a computer using a custom-made graphical user interface. We use a computer-interfaced photodiode (PM16-144,\,Thorlabs) for real-time monitoring of the transmitted power during the alignment process. The actual optical losses were measured using a separate handheld fiber-coupled photodiode (PM20,\,Thorlabs) in order to mitigate errors due to fiber-insertion losses.

\section{Cooldown}
\label{cooldown_section}
\noindent Before the cooldown, the U-bench, together with the filter, was aligned \textit{ex~situ} at room temperature under atmospheric pressure. This involved, in particular, the hand-adjustment of the $xy$-positions of both collimating lenses. Over several test runs, between which the U-bench was misaligned down to unmeasurable transmitted power, we have managed to realign the system to minimize optical losses within the range of 1.55--2.6\,dB (transmission coefficient: $T\approx55$--70\%) under ambient environment. We want to point out that these numbers include possible fiber-coupler losses, and thus they should be considered as the conservative upper limit for the actual U-bench losses. 

\begin{figure*}
    \centering
    \includegraphics[width=0.95\linewidth]{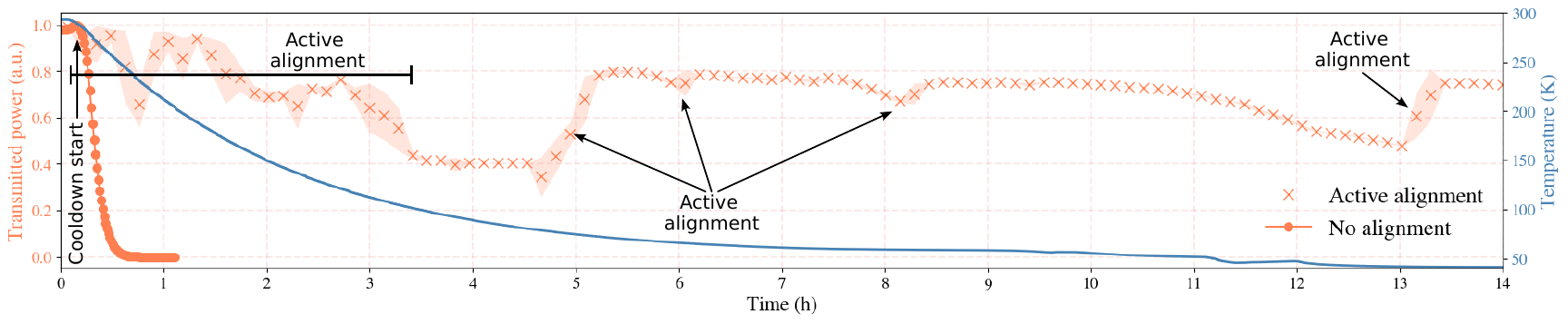}
    \caption{The transmitted power with respect to the initially measured transmission coefficient (we have reached absolute values in the range of $T\approx55$--70\% for several repeated initial alignments) at room temperature (left axis) as a function of elapsed time since the start of the cooldown. The recoded data during the active alignment was sampled at 10\,s intervals, which we have further averaged in the figure. The shaded region corresponds to the standard deviation associated with the average.  The crosses refer to a cooldown during which the U-bench was actively realigned in order to maintain measurable transmission. The obtained transmission coefficient at the end of the cooldown at 40\,K was $T\approx 38\%$. The dots are associated with a cooldown where the U-bench was not realigned after the beginning of the cooldown, resulting in increasing losses over time due to thermal constaction of its components. The corresponding temperatures are illustrated on the right axis.}
    \label{cooldown_fig}
\end{figure*}

After closing the system, we observed an exponential decay in the transmitted power as the vacuum can was pumped down to 1\,mbar. We attribute this to the change of the refractive index of the medium, which alters the focal point of the lens and transmission window of the filter. Regardless, the transmission could be promptly returned back to the ambient level obtained previously by realigning the U-bench \textit{in~situ} with the piezo motors. Upon starting the cooldown by turning on the pulse tube of the dilution refrigerator, we noticed, rather surprisingly, that vibrations originating from the pulse tube or the gas handling system of the cryostat did not have any significant effect on the U-bench performance, despite the absence of any additional implementation of vibration damping. Once the temperature of the 40\,K--flange of the cryostat, and the U-bench, started decreasing at a rate of $\sim 1.5\,$K/min, we observed in real-time how the transmitted power degraded at a steady rate. This must be due to thermal contraction of the U-bench filter's components, gradually misaligning the U-bench until no measurable power is transmitted. In order to keep the transmitted power at a measurable level, which is required for having a feedback loop in the manual alignment process, we actively realigned the U-bench \textit{in~situ} while the system was cooling down. The transmitted power stabilized considerably below 80\,K due to decreasing coefficient of thermal expansion and refrigeration power towards lower temperatures. The temporal evolution of the transmission coefficients during active alignment and passive (alignment-free) cooldowns, together with and associated temperature at the 40\,K-flange of the cryostat, are illustrated in Fig.~\ref{cooldown_fig}. The necessity of active alignment is apparent as the transmission coefficient for passive cooldown drops rapidly with temperature, reaching noise floor at around 240\,K. During the active alignment we noticed that the transmitted power can be maintained by controlling mainly two parameters: \textit{i)} the distance between the fiber tip and collimating lens (focus) of the emitting stage and \textit{ii)} rotating the filter with respect to the beam axis. This suggests that the main temperature dependent loss mechanisms are related to the increased divergence of the emitted beam and blueshift of the filter's transmission window. These observations were done repeatedly over several cooldowns.

At base temperature of 40\,K, the filter bench losses were, again repeatedly, measured to be significantly higher when compared with the 300\,K alignment. For the cooldown illustrated in Fig.~\ref{cooldown_fig}, around 4.2\,dB loss ($T=38\%$) was measured at base temperature, that is approximately 75\% of the maximum transmission observed at room temperature. This is despite our considerable efforts in restoring it back to the ambient level during the cooldown and while stable at the base temperature. The significant difference between the minimum losses measured at 300\,K and 40\,K, respectively, suggest a fundamental limitation in our U-bench design. A possible explanation for this would be the inability for $in~situ$ realignment of the $xy$-positions of the collimating lenses, that are the only four degrees of freedom which are currently out of our reach. Regardless, we consider the achieved 4.2\,dB loss to be rather small for the 260\,K change in temperature, particularly considering the improvement achieved compared with previous work (Fig.~\ref{NEP_and_S_Fig_ab}(a)).

\section{Conclusions}
\noindent We have presented a prototype design for a fiber-optic U-bench ultra-narrow bandpass filter that can be aligned \textit{in~situ} in a cryogenic environment in order to minimize optical losses down to 4.2\,dB, corresponding to 38\% transmission, at 40\,K temperature. However, we have repeatedly measured significantly lower losses at room temperature, down to 1.55\,dB, suggesting a fundamental temperature dependent limitation in our U-bench design that is currently out of our control. Recognizing the limiting feature and subsequently acquiring control over it would thus result in significant further mitigation of losses. A plausible explanation for the increased low temperature losses would be the currently existing inability for \textit{in situ} alignment of positions of the collimating lenses in the plane perpendicular to the beam axis.

The herein used filter has a transmission window of $\Delta \lambda \approx 2\,$nm at a center wavelength of $\lambda=1064\,$nm, which projects to $\dot{n}_\mathrm{b}\sim 10^{-8}\,$Hz of 300\,K black-body background (Fig.~\ref{Ndot-dlam_Fig}). At this level, intrinsic background sources start to limit the dark count rate corresponding to 1064\,nm triggers to $\sim 10^{-6}\,$Hz range~\cite{Meyer2024first}. The implemented U-bench filter shows great potential for improving the sensitivity of rare-event searches, such as Any Light Particle Search~II (ALPS~II) looking for axions and axion-like particles \cite{Brotherton2025any, Spector2026any}. For 1550\,nm wavelength, for which black-body background is limited to $n_\mathrm{b}\sim 10^{-3}\,$Hz, Fig.~\ref{NEP_and_S_Fig_ab}(a) suggests that $\mathrm{NEP}\approx 10^{-20}\,\mathrm{W}\,\mathrm{Hz}^{-1/2}$ is within reach. Future work on the topic includes the full integration of the U-bench filter with our TES module, as illustrated in Fig.~\ref{filter_bench_schematics_Fig_ab}(b). We are also investigating cost effective methods for achieving \textit{in~situ} control of the $xy$-positioners of the collimating lenses, which we expect to further decrease the losses at the 40\,K base temperature.

\section{Acknowledgments}
\noindent The authors acknowledge support from the European Research Council (ERC) under the European Union’s Horizon 2020 research and innovation program Grant agreement No. 948689 (AxionDM). M. M. and J. R. R. also acknowledge support from the Independent Research Fund Denmark (Grant No. 10.46540/5281-00207B). We are grateful to DESY and the ALPS II collaboration for providing the infrastructure and support at DESY in Hamburg. We acknowledge the support of the Deutsche Forschungsgemeinschaft through project grant WI 1643/2-1 and EXC 2121 “Quantum Universe” – 390833306. The authors sincerely thank SDU technician Lars Brændegaard Hansen for designing and constructing the base plate for the optical U-bench, as well as the necessary adapters for mounting the fiber ends, lenses and filter within the setup.

\bibliographystyle{unsrt}
\bibliography{bibliography.bib}

@article{Manenti2024dark,
  title={Dark counts in optical superconducting transition-edge sensors for rare-event searches},
  author={Manenti, Laura and Pepe, Carlo and Sarnoff, Isaac and Ibrayev, Tengiz and Oikonomou, Panagiotis and Knyazev, Artem and Monticone, Eugenio and Garrone, Hobey and Alder, Fiona and Fawwaz, Osama and others},
  journal={Physical Review Applied},
  volume={22},
  number={2},
  pages={024051},
  year={2024},
  publisher={APS}
}

@article{Meyer2024first,
  title={A first application of machine and deep learning for background rejection in the {ALPS II} {TES} detector},
  author={Meyer, Manuel and Isleif, Katharina and Januschek, Friederike and Lindner, Axel and Othman, Gulden and Rubiera Gimeno, Jos{\'e} Alejandro and Schwemmbauer, Christina and Schott, Matthias and Shah, Rikhav and ALPS Collaboration},
  journal={Annalen der Physik},
  volume={536},
  number={1},
  pages={2200545},
  year={2024},
  publisher={Wiley Online Library}
}

@article{Gimeno2024tes,
  title={A {TES} system for {ALPS II} - {S}tatus and Prospects},
  author={Gimeno, Jos{\'e} Alejandro Rubiera and Januschek, Friederike and Isleif, Katharina Sophie and Lindner, Axel and Meyer, Manuel and Othman, Gulden and Schwemmbauer, Christina and Shah, Rikhav},
  journal={Proceedings of Science},
  volume={449},
  pages={567},
  year={2024},
  publisher={SISSA}
}

@article{Rubiera2025simulation,
  title={Simulation and measurement of blackbody radiation background in a transition edge sensor},
  author={Rubiera Gimeno, Jos{\'e} Alejandro and Januschek, Friederike and Lindner, Axel and Schwemmbauer, Christina and Isleif, Katharina-Sophie and Meyer, Manuel and Rivasto, Elmeri and Othman, Gulden and Shah, Rikhav},
  journal={Physical Review D},
  volume={112},
  number={3},
  pages={032001},
  year={2025},
  publisher={APS}
}

@inproceedings{Miller2007superconducting,
  title={Superconducting photon number resolving detectors: Performance and promise},
  author={Miller, Aaron J and Lita, Adriana and Rosenberg, Danna and Gruber, Stephen and Nam, S},
  booktitle={Proc. 8th Int. Conf. Quantum Communication, Measurement and Computing (QCMC’06)},
  pages={445--450},
  year={2007}
}

@article{Hadfield2009single,
  title={Single-photon detectors for optical quantum information applications},
  author={Hadfield, Robert H},
  journal={Nature photonics},
  volume={3},
  number={12},
  pages={696--705},
  year={2009},
  publisher={Nature Publishing Group UK London}
}

@article{Morozov2021superconducting,
  title={Superconducting photon detectors},
  author={Morozov, Dmitry V and Casaburi, Alessandro and Hadfield, Robert H},
  journal={Contemporary Physics},
  volume={62},
  number={2},
  pages={69--91},
  year={2021},
  publisher={Taylor \& Francis}
}

@article{Fukuda2011titanium,
  title={Titanium-based transition-edge photon number resolving detector with 98\% detection efficiency with index-matched small-gap fiber coupling},
  author={Fukuda, Daiji and Fujii, Go and Numata, Takayuki and Amemiya, Kuniaki and Yoshizawa, Akio and Tsuchida, Hidemi and Fujino, Hidetoshi and Ishii, Hiroyuki and Itatani, Taro and Inoue, Shuichiro and others},
  journal={Optics express},
  volume={19},
  number={2},
  pages={870--875},
  year={2011},
  publisher={Optical Society of America}
}

@article{Lita2008counting,
  title={Counting near-infrared single-photons with 95\% efficiency},
  author={Lita, Adriana E and Miller, Aaron J and Nam, Sae Woo},
  journal={Optics express},
  volume={16},
  number={5},
  pages={3032--3040},
  year={2008},
  publisher={Optical Society of America}
}

@book{Donati2021photodetectors,
  title={Photodetectors: devices, circuits and applications},
  author={Donati, Silvano},
  year={2021},
  publisher={John Wiley \& Sons}
}

@article{Niwa2017few,
  title={Few-photon color imaging using energy-dispersive superconducting transition-edge sensor spectrometry},
  author={Niwa, Kazuki and Numata, Takayuki and Hattori, Kaori and Fukuda, Daiji},
  journal={Scientific Reports},
  volume={7},
  number={1},
  pages={45660},
  year={2017},
  publisher={Nature Publishing Group UK London}
}

@article{Yang2014superconducting,
  title={Superconducting nanowire single photon detector with on-chip bandpass filter},
  author={Yang, Xiaoyan and Li, Hao and Zhang, Weijun and You, Lixing and Zhang, Lu and Liu, Xiaoyu and Wang, Zhen and Peng, Wei and Xie, Xiaoming and Jiang, Mianheng},
  journal={Optics express},
  volume={22},
  number={13},
  pages={16267--16272},
  year={2014},
  publisher={Optical Society of America}
}

@article{Yamashita2010temperature,
  title={Temperature dependent performances of superconducting nanowire single-photon detectors in an ultralow-temperature region},
  author={Yamashita, Taro and Miki, Shigehito and Qiu, Wei and Fujiwara, Mikio and Sasaki, Masahide and Wang, Zhen},
  journal={Applied physics express},
  volume={3},
  number={10},
  pages={102502},
  year={2010},
  publisher={IOP Publishing}
}

@article{Eisaman2011invited,
  title={Invited review article: Single-photon sources and detectors},
  author={Eisaman, Matthew D and Fan, Jingyun and Migdall, Alan and Polyakov, Sergey V},
  journal={Review of scientific instruments},
  volume={82},
  number={7},
  year={2011},
  publisher={AIP Publishing}
}

@book{Schroeder2020introduction,
  title={An introduction to thermal physics},
  author={Schroeder, Daniel V},
  year={2020},
  publisher={Oxford University Press}
}

@article{De2024transition,
  title={Transition edge sensors: Physics and applications},
  author={De Lucia, Mario and Dal Bo, Paolo and Di Giorgi, Eugenia and Lari, Tommaso and Puglia, Claudio and Paolucci, Federico},
  journal={Instruments},
  volume={8},
  number={4},
  pages={47},
  year={2024},
  publisher={MDPI}
}

@article{Shibata2013superconducting,
  title={Superconducting nanowire single-photon detector with ultralow dark count rate using cold optical filters},
  author={Shibata, Hiroyuki and Shimizu, Kaoru and Takesue, Hiroki and Tokura, Yasuhiro},
  journal={Applied Physics Express},
  volume={6},
  number={7},
  pages={072801},
  year={2013}
}

@article{Shibata2014quantum,
  title={Quantum key distribution over a 72 dB channel loss using ultralow dark count superconducting single-photon detectors},
  author={Shibata, Hiroyuki and Honjo, Toshimori and Shimizu, Kaoru},
  journal={Optics letters},
  volume={39},
  number={17},
  pages={5078--5081},
  year={2014},
  publisher={Optical Society of America}
}

@article{Shibata2015ultimate,
  title={Ultimate low system dark-count rate for superconducting nanowire single-photon detector},
  author={Shibata, Hiroyuki and Shimizu, Kaoru and Takesue, Hiroki and Tokura, Yasuhiro},
  journal={Optics letters},
  volume={40},
  number={14},
  pages={3428--3431},
  year={2015},
  publisher={Optical Society of America}
}

@article{Shibata2016snspd,
  title={{SNSPD} with ultimate low system dark count rate using various cold filters},
  author={Shibata, Hiroyuki and Fukao, Kentaro and Kirigane, Naoto and Karimoto, Shinichi and Yamamoto, Hideki},
  journal={IEEE Transactions on Applied Superconductivity},
  volume={27},
  number={4},
  pages={1--4},
  year={2016},
  publisher={IEEE}
}

@article{Hummatov2023fast,
  title={Fast transition-edge sensors suitable for photonic quantum computing},
  author={Hummatov, Ruslan and Lita, Adriana E and Farrahi, Tannaz and Otrooshi, Negar and Fayer, Samuel and Collins, Matthew J and Durkin, Malcolm and Bennett, Douglas and Ullom, Joel and Mirin, Richard P and others},
  journal={Journal of Applied Physics},
  volume={133},
  number={23},
  year={2023},
  publisher={AIP Publishing}
}

@article{Li2025tes,
  title={A {TES} single-photon detector demonstrating high-energy-resolution, high efficiency, and ultra-low dark counts near $0.9\,{T}_\mathrm{c}$},
  author={Li, PZ and Zhang, W and Feng, ZF and Wang, Z and Ma, QX and Zhong, JQ and Miao, W and Yao, QJ and Li, J and Shi, SC},
  journal={APL Photonics},
  volume={10},
  number={12},
  year={2025},
  publisher={AIP Publishing}
}

@article{Rivasto2026binary,
  title={Binary classification of signal and background triggers of a transition edge sensor using convolutional neural networks},
  author={Rivasto, Elmeri and Isleif, Katharina-Sophie and Januschek, Friederike and Lindner, Axel and Meyer, Manuel and Othman, Gulden and Rubiera Gimeno, Jos{\'e} Alejandro and Schwemmbauer, Christina},
  journal={Scientific Reports},
  volume={16},
  number={1},
  pages={3389},
  year={2026},
  publisher={Nature Publishing Group UK London}
}

@article{Brotherton2025any,
  title={Any Light Particle Searches with {ALPS} {II}: {F}irst science results},
  author={Brotherton, Daniel C and Bush, Zachary R and Croatto, Sandy and Egge, Jacob and Ejlli, Aldo and Fr{\"a}drich, Henry and Gleason, Joe and Grote, Hartmut and Hallal, Ayman and Hartman, Michael T and others},
  journal={arXiv preprint arXiv:2512.14110},
  year={2025}
}

@article{Spector2026any,
  title={Any Light Particle Searches with {ALPS} {II}: Description of the first science campaign},
  author={Spector, Aaron D and Brotherton, Daniel C and Hallal, Ayman and Fr{\"a}drich, Henry and Egge, Jacob and Wei, Li-Wei and Kozlowski, Todd and Karan, Kanioar and Bush, Zachary R and Ejlli, Aldo and others},
  journal={arXiv preprint arXiv:2601.18684},
  year={2026}
}

@article{Shah2022characterising,
  title={Characterising a single-photon detector for {ALPS} {II}},
  author={Shah, Rikhav and Isleif, Katharina-Sophie and Januschek, Friederike and Lindner, Axel and Schott, Matthias},
  journal={Journal of Low Temperature Physics},
  volume={209},
  number={3},
  pages={355--362},
  year={2022},
  publisher={Springer}
}

@article{Meyer2026characterization,
  title={Characterization of a Two-Channel Optical and Near-infrared Transition Edge Sensor System for Rare-Event Searches},
  author={Meyer, Manuel and Isleif, Katharina-Sophie and Januschek, Friederike and Lindner, Axel and Othman, Gulden and Rivasto, Elmeri and Gimeno, Jose Alejandro Rubiera and Schwemmbauer, Christina},
  journal={arXiv preprint arXiv:2605.07423},
  year={2026}
}

@article{Shah2022TES,
  author = "Shah, Rikhav  and  Isleif, Katharina-Sophie  and  Januschek, Friederike  and  Lindner, Axel  and  Schott, Matthias",
  title = {{TE}S Detector for {ALPS} {II}},
  doi = "10.22323/1.398.0801",
  journal = "PoS",
  year = 2022,
  volume = "EPS-HEP2021",
  pages = "801"
}

@article{Hu2020detecting,
  title={Detecting single infrared photons toward optimal system detection efficiency},
  author={Hu, Peng and Li, Hao and You, Lixing and Wang, Heqing and Xiao, You and Huang, Jia and Yang, Xiaoyan and Zhang, Weijun and Wang, Zhen and Xie, Xiaoming},
  journal={Optics Express},
  volume={28},
  number={24},
  pages={36884--36891},
  year={2020},
  publisher={Optical Society of America}
}

\end{document}